\documentclass[showkeys,showpacs,superscriptaddress,nofootinbib]{revtex4-2}
\usepackage{amsmath}
\usepackage{amssymb}
\usepackage{dutchcal}
\usepackage{graphicx}
\usepackage[normalem]{ulem}
\usepackage{xcolor}
\usepackage{float}
\usepackage{slashed}

\newcommand{\be}{\begin{equation}}
\newcommand{\ee}{\end{equation}}

\allowdisplaybreaks

\begin{document}

\title{Non-Abelian monopoles in Einstein--scalar--Gauss--Bonnet gravity
}
\author{
Vladimir Dzhunushaliev
}
\email{v.dzhunushaliev@gmail.com}
\affiliation{
Department of Theoretical and Nuclear Physics,  Farabi University, Almaty 050040, Kazakhstan
}
\affiliation{
Institute for Experimental and Theoretical Physics, Farabi University, Almaty 050040, Kazakhstan
}
\affiliation{Academician J.~Jeenbaev Institute of Physics of the NAS of the Kyrgyz Republic, 265 a, Chui Street, Bishkek 720071, Kyrgyzstan}

\author{Vladimir Folomeev}
\email{vfolomeev@mail.ru}


\affiliation{Academician J.~Jeenbaev Institute of Physics of the NAS of the Kyrgyz Republic, 265 a, Chui Street, Bishkek 720071, Kyrgyzstan}

\begin{abstract}
We study static, spherically symmetric self-gravitating non-Abelian 't Hooft--Polyakov monopoles in Einstein--scalar--Gauss--Bonnet gravity 
for two classes of scalar--Gauss--Bonnet coupling functions. We demonstrate that the branch structure of the monopole solutions depends sensitively 
on the choice of the coupling function. For the polynomial coupling, we show that, for sufficiently large values of the coupling parameter, 
the principal part of the field equations becomes locally degenerate at a critical radius, leading to a critical solution that signals a geometric phase transition. 
By contrast, the exponential coupling acts as a natural regulator, preventing the degeneration of the principal part and preserving smooth field profiles. 
Finally, we discuss a possible physical interpretation of the critical solution, in which spacetime separates into an inner region 
where quantum-gravitational effects may become important and an outer region that remains well described by classical general relativity.
\end{abstract}

\pacs{}

\keywords{Non-Abelian monopole, Einstein--scalar--Gauss--Bonnet gravity, geometric phase transition }
\date{\today}

\maketitle 

\section{Introduction}

General relativity (GR) has been remarkably successful in describing gravitational phenomena over a wide range of scales. 
Its predictions have been confirmed with high precision in Solar System experiments, binary pulsar observations, and, more recently, 
by gravitational-wave detections from compact binary mergers. Despite this outstanding observational success, GR is generally regarded 
as an effective classical theory rather than a complete description of gravity. In particular, it is perturbatively nonrenormalizable 
and therefore cannot be straightforwardly quantized within the framework of conventional quantum field theory. 
Furthermore, several fundamental problems in modern cosmology and astrophysics, including the nature of dark matter, dark energy, 
the initial singularity, and the late-time accelerated expansion of the Universe, suggest that Einstein's theory may require modifications in the ultraviolet or infrared regimes.

These motivations have stimulated extensive research on modified theories of gravity. Among the numerous proposals, 
Einstein--scalar--Gauss--Bonnet (EsGB) gravity occupies a particularly distinguished position. The theory extends the Einstein--Hilbert 
action by coupling a scalar field to the Gauss--Bonnet invariant, thereby introducing higher-curvature corrections while preserving 
second-order field equations and avoiding the Ostrogradsky instability. Such couplings naturally arise as leading-order corrections 
in the low-energy effective action of heterotic string theory~\cite{Gross:1986mw}
and other ultraviolet completions of gravity. 
Consequently, EsGB gravity provides a theoretically well-motivated framework for investigating deviations from GR in regimes of strong spacetime curvature.

During the last decade, EsGB gravity has attracted considerable attention in a broad range of applications. 
The theory admits black-hole solutions with nontrivial scalar hair~\cite{Doneva:2022ewd}, thereby circumventing classical no-hair theorems 
through spontaneous scalarization or nonminimal scalar-curvature couplings. It also supports neutron stars \cite{Doneva:2017duq}
and other ultracompact objects \cite{Kleihaus:2019rbg} with properties that can significantly differ from those predicted by GR, offering 
potential observational signatures in gravitational-wave astronomy and electromagnetic observations. 
In addition, EsGB gravity has been extensively employed in studies of traversable wormholes without exotic matter \cite{Kanti:2011jz,Kanti:2011yv,Antoniou:2019awm}, 
early-Universe inflation \cite{Kanti:2015pda,Odintsov:2018zhw}, nonsingular cosmological models \cite{Antoniadis:1993jc,Kanti:1998jd}, late-time cosmic acceleration \cite{Nojiri:2005vv,Cognola:2006eg,Nojiri:2010wj}, 
and various aspects of strong-field gravitational physics. 
These diverse applications demonstrate that EsGB gravity provides a versatile framework for exploring both astrophysical and cosmological consequences of higher-curvature corrections.

Topological solitons constitute another important class of nonlinear field configurations whose properties may be substantially 
altered by modifications of the gravitational interaction. Among them, the 't Hooft--Polyakov monopole remains one of the most 
fundamental solutions of spontaneously broken non-Abelian gauge theories~\cite{tHooft:1974kcl,Polyakov:1974ek}. In Einstein gravity, gravitating monopoles 
have been investigated in detail and exhibit a rich spectrum of regular~\cite{Breitenlohner:1991aa,Breitenlohner:1994di,Kleihaus:1997ic,Volkov:1998cc,Hartmann:2001ic} 
and black-hole solutions together with critical phenomena 
associated with gravitational collapse~\cite{Lee:1991vy,Volkov:1989fi,Volkov:1990sva,Bizon:1990sr,Brihaye:1999nn}. 
In contrast, considerably less attention has been devoted to monopoles in modified theories of gravity, 
especially in theories containing higher-curvature corrections~\cite{Edery:2018jyp}. Only a limited number of works have addressed gravitating monopoles 
in frameworks such as $f(R)$ gravity~\cite{Dzhunushaliev:2026slt} or scalar-tensor theories~\cite{Hartmann:2002ez}, while systematic studies in Einstein--scalar--Gauss--Bonnet gravity remain largely absent.

The purpose of the present work is to investigate self-gravitating 't Hooft--Polyakov monopoles in Einstein--scalar--Gauss--Bonnet gravity. 
We derive the corresponding field equations, construct globally regular monopole nodeless solutions numerically, and analyze 
the influence of the Gauss--Bonnet coupling on the structure of the monopole, including its mass
and domain of existence. Our results provide some new insights into the interplay between non-Abelian gauge fields
and higher-curvature corrections to Einstein gravity, thereby extending the current understanding of topological solitons in modified gravitational theories.

\section{The model}
We work within EsGB gravity and consider SU(2) Yang-Mills fields $A^a_\mu$ and scalar (triplet) fields $\phi^a$   with the action
[we use signature $(+,-,-,-)$ and units $c=\hbar=1$]
\begin{equation}
S=\int d^4x~\sqrt{-g}\left\{
-\frac{R}{16 \pi G} + f(\Phi) R_{\text{GB}}^2 + L_m
    \, \right\},
\label{action}
\end{equation}
where $G$ is the Newtonian gravitational constant, $g$
denotes the determinant of the metric tensor, 
$\Phi=\phi^a\phi^a$, 
$
R_{\text{GB}}^2=R^2-4 R_{\mu\nu}R^{\mu\nu}+R_{\mu\nu\rho\sigma}R^{\mu\nu\rho\sigma}
$ is the Gauss-Bonnet invariant,
and the matter field Lagrangian is
$$
L_m=
- \frac{1}{4} F_{\mu \nu}^a F^{a\mu \nu}   + \frac{1}{2} D_\mu \phi^a D^\mu \phi^a  -
	\frac{\lambda}{4} \left( \Phi - \phi_0^2 \right)^2 .
$$
Here 
$
	F^a_{\mu \nu} = \partial_\mu A^a_\nu - \partial_\nu A^a_\mu +
	e \epsilon^{a b c} A^b_\mu A^c_\nu
$ is the field strength tensor for the Yang-Mills field, where $\epsilon^{a b c}$ are the SU(2) structure constants (the completely antisymmetric Levi-Civita symbol), $e$ is the gauge coupling constant,
$a, b, c = 1, 2, 3$ are SU(2) color indices, $\mu, \nu = 0, 1, 2, 3$ are spacetime indices; $D_\mu \phi^a = \partial_\mu \phi^a + e \epsilon^{a b c} A_\mu^b \phi^c$. 
The action also contains a Higgs vacuum expectation value $\phi_0$ and the scalar self-interaction constant $\lambda$ which
defines the mass of the Higgs field, $M_s=\sqrt \lambda \phi_0$. The gauge field becomes
massive due to the coupling with the scalar field, $M_v=e\phi_0$.

By varying the action \eqref{action} with respect to $g_{\mu\nu}$, we have the gravitational equations
\begin{equation}
 R^\mu_\nu-\frac{1}{2} \delta^\mu_\nu R=
8\pi G\left[T^\mu_{\nu \text{(YMH)}}+T^\mu_{\nu \text{(GB)}}
\right] ,
\label{grav_eqs}
\end{equation}
where 
$$
T^\mu_{\nu \text{(YMH)}}=-F^{a\mu\alpha}F^a_{\nu\alpha}+D^\mu\phi^a D_\nu\phi^a-
\delta_\nu^\mu L_m
$$
and
\begin{equation*}
\begin{aligned}
T_{\mu \nu }^{(GB)}=&-4\left(\nabla _{\mu }\nabla _{\nu }f\right)R+4g_{\mu \nu }\left(\Box f\right)R-8\left(\Box f\right)R_{\mu \nu } 
+8\left(\nabla ^{\rho }\nabla _{\mu }f\right)R_{\nu \rho }+8\left(\nabla ^{\rho }\nabla _{\nu }f\right)R_{\mu \rho }\\ 
&-8g_{\mu \nu }\left(\nabla ^{\rho }\nabla ^{\sigma }f\right)R_{\rho \sigma } +8\left(\nabla ^{\rho }\nabla ^{\sigma }f\right)R_{\mu \rho \nu \sigma } 
\end{aligned}
\label{EMT_GB}
\end{equation*}
with
$$
\nabla_{\mu} \nabla_{\nu} f =f_{\Phi\Phi} (\nabla_\mu\Phi) (\nabla_\nu\Phi)+ f_\Phi \left(\partial_\mu \partial_\nu \Phi - \Gamma^\lambda_{\mu\nu} \partial_\lambda \Phi\right),
$$
where
 $ f_\Phi\equiv df/d\Phi$.

In turn, the matter fields equations are derived by varying \eqref{action} with respect to $ A^a_\mu$ and $\phi^a$:
\be
\begin{split}
&D_\nu F^{a \nu \mu}\equiv\frac{1}{\sqrt{-g}}\frac{\partial}{\partial x^\nu}\left(\sqrt{-g}F^{a\nu\mu}\right)+e \epsilon^{a b c}A^b_\nu F^{c\nu\mu} 
 =  -e \epsilon^{abc} \phi^b D^\mu \phi^c     \, , \\
&D_\mu D^\mu \phi^a + \lambda \phi^a \left(\phi^b \phi^b - \phi_0^2 \right)   - 2 \phi^a f_\Phi R_{\text{GB}}^2  =  0  .
\label{field_eqs}
\end{split}
\ee

In what follows, for the coupling function $f(\Phi)$, we employ a polynomial 
\begin{equation}
f(\Phi)=\gamma \left(\Phi-\phi_0^2\right)^2 
\label{pow_coup}
\end{equation}
and an exponential
\begin{equation}
f(\Phi)= \gamma\left[1-e^{-\delta\left(\Phi-\phi_0^2\right)^2 } \right]
\label{exp_coup}
\end{equation}
functions, where  $\gamma$ and $\delta$ are some constants.
Such functions preserve the GR vacuum far from the monopole, i.e., they localize the modification of the gravitational sector within the monopole core. 
Furthermore, as shown below, they provide a qualitatively different structure of the solution branches.

\section{Equations and solutions}

Working within the model described above, in this section we present the general spherically symmetric equations and solve them numerically
for various values of the system parameters.

\subsection{The Ansatz}

For the gauge and Higgs fields, we employ the usual static spherically symmetric
hedgehog Ansatz \cite{tHooft:1974kcl,Polyakov:1974ek}
\be
A_0^a=0,\quad A_i^a= \varepsilon_{aik} \frac{r^k}{er^2} \left[ 1 - W(r) \right],\qquad
\phi^a=\frac{r^a}{e r}H(r)\,.
\label{fields-boson}
\ee 

For the line element, we adopt Schwarzschild-like coordinates
\begin{equation}
    ds^2 = e^{2\nu(r)} dt^2 - e^{2\lambda(r)} dr^2- r^2 \left(d\theta^2 + \sin^2 \theta d\varphi^2\right) .
\label{metric_sph}
\end{equation}
The metric function $\nu(r)$ can be rewritten as $e^{2\nu(r)}=1-\frac{2 G \mu(r)}{r}$ with the mass function $\mu(r)$;  the ADM mass of the
configuration is defined as $M=\mu(\infty)$.

\subsection{Equations and boundary conditions}

Substitution of
the Ansatz \eqref{fields-boson} and \eqref{metric_sph}
into the general set of equations \eqref{grav_eqs} and \eqref{field_eqs} yields the following
set of five coupled ordinary differential equations for the functions $W,H,\nu$, and $\lambda$
(here the prime denotes differentiation  with respect to the radial coordinate, $\nu ^\prime = \frac{d \nu}{dx}$, etc.):
\begin{align}
    & W^{\prime\prime}+ \left(\nu^\prime - \lambda^\prime\right)W^\prime +e^{2\lambda}\left[\frac{1}{x^2}\left(1-W^2\right)-H^2\right]W    = 0 \, ,
\label{YM_eqs}\\
& H^{\prime\prime}+\left(\frac{2}{x}+\nu^\prime - \lambda^\prime\right)H^{\prime}-\frac{4\alpha^2}{x^2}f_\Phi H \left[\left(1-e^{-2\lambda}\right)\left(\nu^{\prime\prime}+\nu^{\prime 2}\right)
-\left(1-3 e^{-2\lambda}\right)  \nu^{\prime } \lambda^{\prime }\right] \nonumber \\
&+e^{2\lambda}\left[\beta^2\left(1-H^2\right)-\frac{2}{x^2} W^2\right]H = 0\, ,
\label{H_eqn}\\
&\left[1+\frac{4\alpha^4}{x}\left(1-3 e^{-2\lambda}\right) f_\Phi H H^\prime\right]\nu^\prime+\frac{1}{2x}\left(1- e^{2\lambda}\right)\nonumber \\
&+\alpha^2 e^{2\lambda}x\left\{\frac{\beta^2}{4}\left(1-H^2\right)^2 +\frac{H^2 W^2}{x^2}-\frac{1}{2} e^{-2\lambda}H^{\prime 2}+
\frac{1}{2 x^4}\left[\left(1-W^2\right)^2-2 e^{-2\lambda}x^2 W^{\prime 2}\right]
\right\} = 0\, ,
\label{Phi1_eqn}\\
&\left[1+\frac{4\alpha^4}{x}\left(1-3 e^{-2\lambda}\right) f_\Phi H H^\prime\right]\lambda^\prime-\frac{1}{2x}\left(1-e^{2\lambda}\right)\nonumber \\
&-\frac{\alpha^2}{2} e^{2\lambda}x\left\{\frac{\beta^2}{2}\left(1-H^2\right)^2+2\frac{H^2 W^2}{x^2}+e^{-2\lambda}H^{\prime 2}+\frac{1}{x^4}\left[\left(1-W^2\right)^2+2 e^{-2\lambda} x^2 W^{\prime 2}\right]
\right\} \nonumber\\
&-\frac{4\alpha^4}{x}f_\Phi\left(1-e^{-2\lambda}\right)\left(H H^{\prime\prime}+H^{\prime 2}\right)
-\frac{2\alpha^6}{x}\left(1-e^{-2\lambda}\right) f_{\Phi\Phi} H^2 H^{\prime 2} = 0\, ,
\label{Lambda_eqn}\\
&\left(1-\frac{8\alpha^4}{x}e^{-2\lambda} f_\Phi H H^\prime\right)\left(\nu^{\prime\prime}+\nu^{\prime 2}\right)-\frac{\lambda^\prime}{x}
+\alpha^2\left[\frac{\beta^2}{2}e^{2\lambda}\left(1-H^2\right)^2-\frac{1}{x^4} e^{2\lambda}\left(1-W^2\right)^2+H^{\prime 2}
\right] \nonumber\\
&+\left[1-4\alpha^6 e^{-2\lambda} f_{\Phi\Phi} H^2 H^{\prime 2}-x \lambda^{\prime}-8 \alpha^4 e^{-2\lambda}f_{\Phi}\left(H H^{\prime\prime}+H^{\prime 2}-3 H H^{\prime }\lambda^\prime\right)
\right]\frac{\nu^\prime}{x} =0 .
\label{Phi2_eqn}
\end{align}
Equations \eqref{YM_eqs}--\eqref{Phi2_eqn} are written in terms of the following dimensionless (tilded) variables and parameters:
a  radial coordinate, $x=e \phi_0 r$,  two rescaled  effective
coupling constants $\alpha^2=4\pi G\phi_0^2\,, \beta^2=\lambda/e^2$,
the scalar field $\tilde H=H/(e \phi_0)$, $\tilde f_{\Phi}=e^2 f_{\Phi}/(\pi G)$, and $\tilde f_{\Phi\Phi}=e^2 f_{\Phi\Phi}/(\pi G)^2$
(to simplify the notation, we omit the tilde from now on).
For numerical calculations, we will use Eqs.~\eqref{YM_eqs}--\eqref{Lambda_eqn}; moreover, in the equation for the scalar field~\eqref{H_eqn}, we will eliminate
 $\nu^{\prime\prime}$ using Eq.~\eqref{Phi2_eqn}, and in Eq.~\eqref{Lambda_eqn}, we will eliminate  $H^{\prime\prime}$ using the resulting equation for $H$.

Since this paper focuses on the influence coming from the modification of Einstein's gravity, 
here we will study the corresponding changes in the total mass of the configurations as the modification of gravity is included.
The dimensionless ADM mass $\tilde M$ can be extracted from the metric function $\nu$ [see after Eq.~\eqref{metric_sph}] as follows:
\begin{equation}
\label{expres_mass}
\tilde M \equiv \frac{e}{4\pi \phi_0}M=\frac{1}{\alpha^2}\lim_{x\to\infty} x^2 e^{2\nu}\partial_x \nu = \frac{c_k}{2\alpha^2}\lim_{\bar x\to 1}\partial_{\bar x} \nu ,
\end{equation}
where  the last expression in the above equation represents the mass in terms of the
compactified coordinate $\bar x$ from Eq.~\eqref{comp_coord}.

The set of mixed-order ODEs \eqref{YM_eqs}--\eqref{Lambda_eqn} is solved numerically subject to  boundary conditions derived from
the asymptotic expansion of the solutions at the boundaries of the integration domain, 
assuming regularity and asymptotic flatness.
 Explicitly, we impose:
\be
\begin{split}
\partial_x \nu(0)=&0, \quad  \lambda(0)=0, \quad W(0)=1,\quad H(0)=0; \nonumber\\
\nu(\infty)=&0, \quad \lambda(\infty)=0, \quad W(\infty)=0,\quad H(\infty)=1 \, . \nonumber
\end{split}
\ee

\subsection{Numerical method}
\label{num_meth}

In order to map the infinite range of the radial variable $x$ 
to a finite interval, we introduce the compactified coordinate~$\bar x$ as follows:
\begin{equation}
     x = c_k\frac{\bar x}{1-\bar x^2} \, ,
\label{comp_coord}
\end{equation}
which maps the infinite region $[0;\infty)$ onto the finite interval $[0; 1]$. Here, $c_k$ is a constant used to adjust grid contraction.
In our calculations, we typically set $c_k=5$, though in some cases 
larger values (up to 8) are chosen to obtain solutions for $\alpha$ approaching critical values  (see below).

The equations are discretized on a grid, and the resulting set of nonlinear algebraic equations is solved using a modified Newton method. 
The underlying linear system is solved with the Intel MKL PARDISO sparse direct solver~\cite{pardiso} 
and the CESDSOL library\footnote{Complex Equations-Simple Domain 
partial differential equations SOLver, a C++ package developed by I.~Perapechka;
see Refs.~\cite{Herdeiro:2019mbz,Herdeiro:2021jgc}.}. 
 Eqs.~\eqref{YM_eqs}--\eqref{Lambda_eqn}
are discretized on a grid of approximately 1000 grid points, with up to 3000 or more points used in certain cases. 
In all cases, the typical errors are on the order of $10^{-4}$.
The package provides an iterative procedure to obtain an exact solution starting from some initial guess configuration. 
For the latter, we use the gravitating monopole configuration found in Ref.~\cite{Breitenlohner:1991aa}.

\subsection{Results of numerical calculations}

\begin{figure}[t!]
\begin{minipage}[t]{.49\linewidth}
        \begin{center}
\includegraphics[width=1.\linewidth]{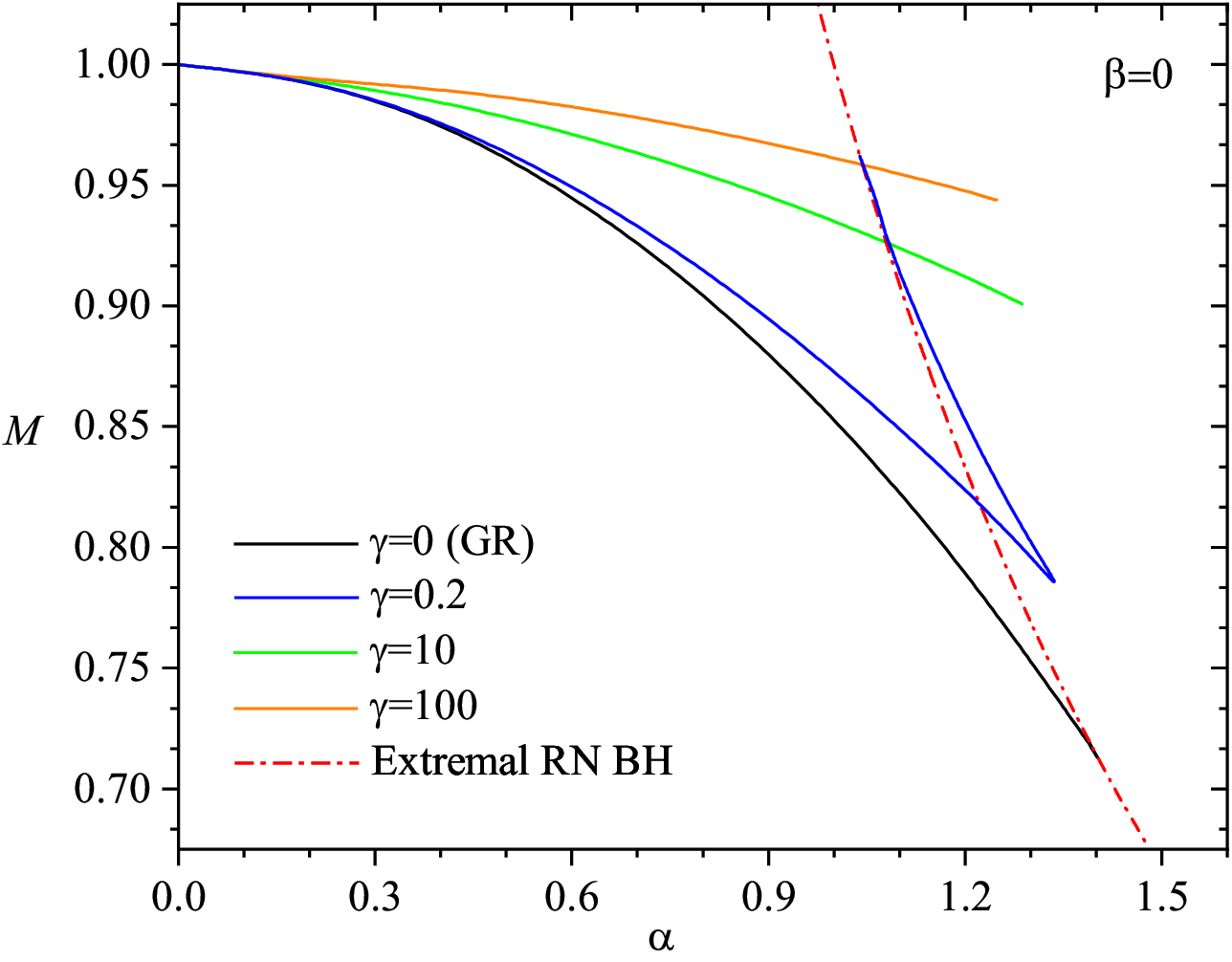}
        \end{center}
\end{minipage}\hfill
\begin{minipage}[t]{.49\linewidth}
        \begin{center}
\includegraphics[width=1.\linewidth]{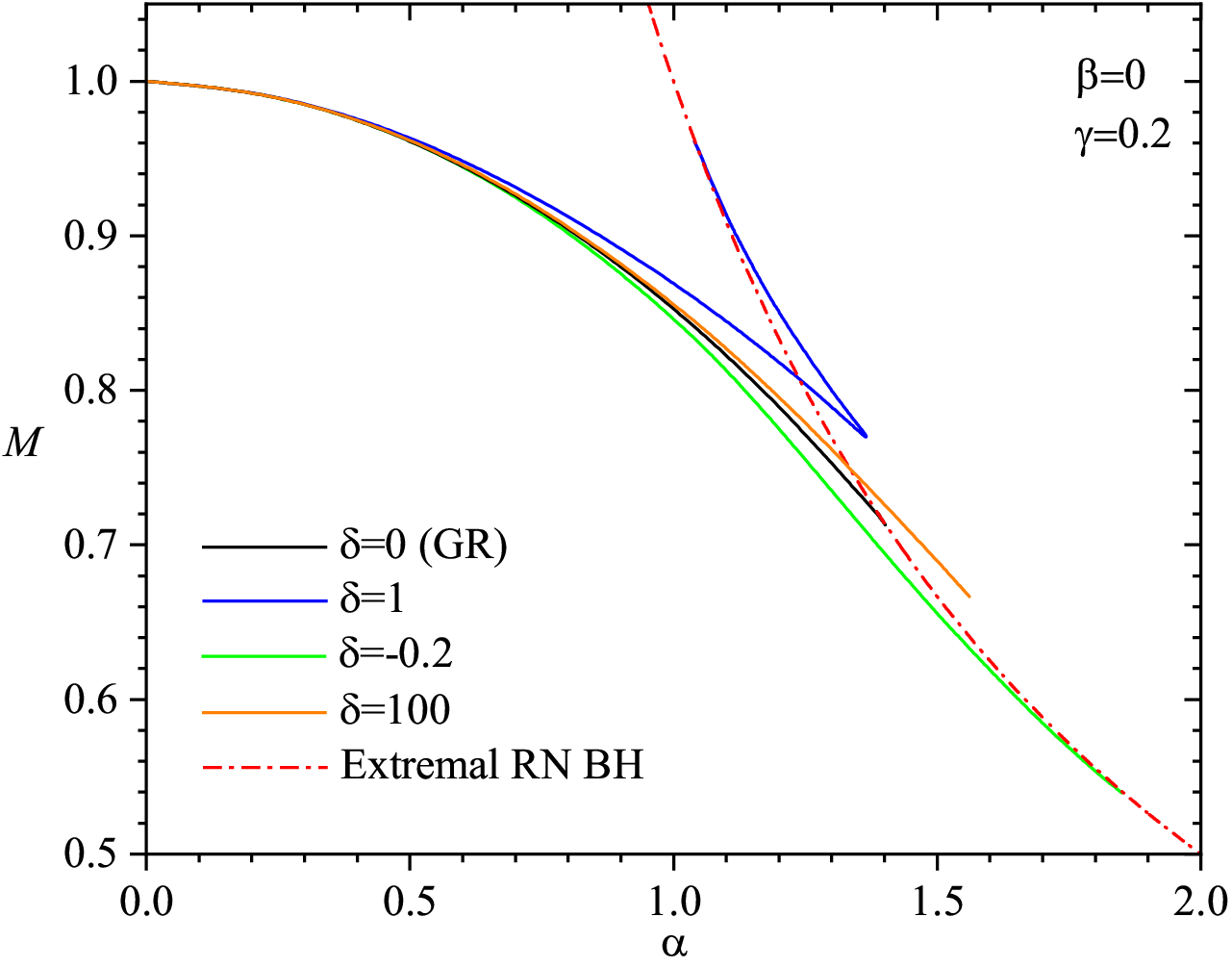}
        \end{center}
\end{minipage}\hfill
\caption{The dependence of the dimensionless ADM mass from Eq.~\eqref{expres_mass} on the effective
gravitational coupling constant $\alpha$ is shown for $\beta=0$ and various $\gamma$ and $\delta$.
The left panel is for the coupling function~\eqref{pow_coup}. The curves with $\gamma\gg 1$ terminate at points $\alpha=\alpha_{\text{max}}^{\text{GB}}$
where the corresponding determinant of the principal-part matrix develops a zero.
The right panel is for the coupling function~\eqref{exp_coup}  with $\gamma=0.2$.
The curve with $\delta=100$ terminates at $\alpha\to \alpha_{\text{max}}^{\text{GB}}\approx\alpha_{\text{cr}}^{\text{GB}}$, 
and the limiting configuration appears to approach a hairy (non-RN) black-hole solution.
For comparison, the mass of an extremal RN black hole of unit charge $\tilde M=1/\alpha$ in GR~\cite{Hartmann:2001ic} is also shown.
}
\label{fig_mass_alpha}
\end{figure}

\begin{figure}[t!]
\begin{center}
\includegraphics[width=1.\linewidth]{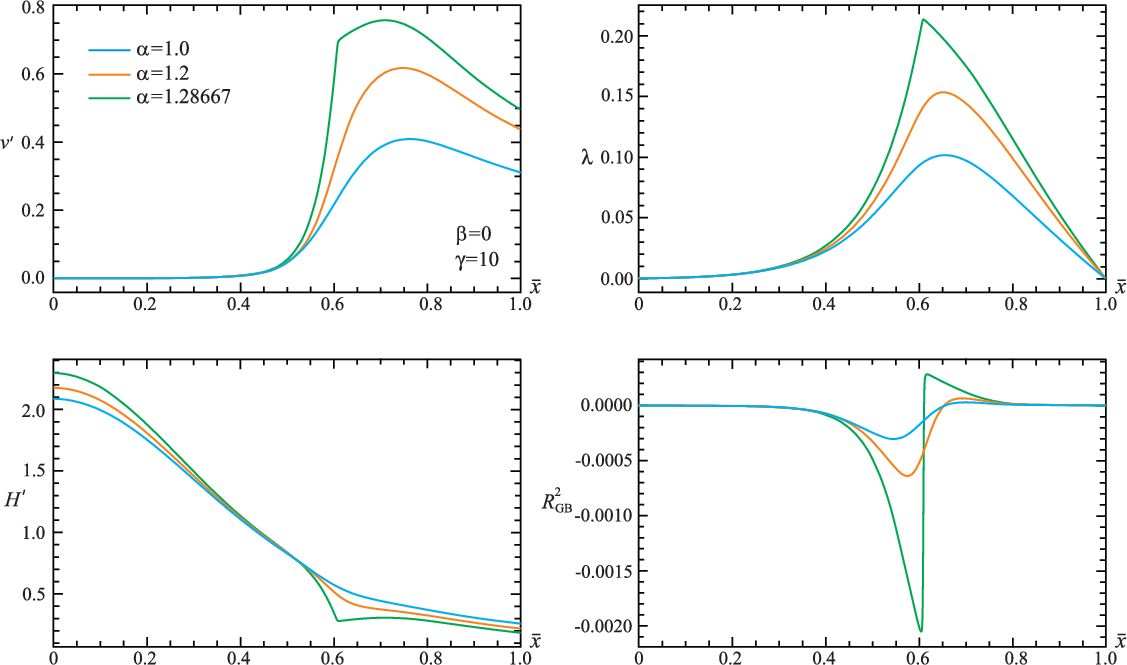}
\end{center}
\vspace{-0.5cm}
\caption{
Evolution of the system with the coupling function \eqref{pow_coup} with $\gamma=10$ and $\beta=0$ with increasing $\alpha$.
A cusp singularity emerges at  $\alpha_{\text{max}}^{\text{GB}}\approx 1.28667$ and $\bar x_{\text{max}}\approx 0.61$. 
The plots are made using the compactified coordinate~$\bar x$ from \eqref{comp_coord} with $c_k=6$. 
(Note that in terms of this coordinate, the derivatives $d H/d \bar{x}$ and $d \nu/d \bar{x}$ do not vanish asymptotically ($\bar{x}\to 1$), whereas in regular coordinates they are zero.)
}
\label{fig_sols}
\end{figure}

\begin{figure}[t!]
\begin{minipage}[t]{.49\linewidth}
        \begin{center}
\includegraphics[width=1.\linewidth]{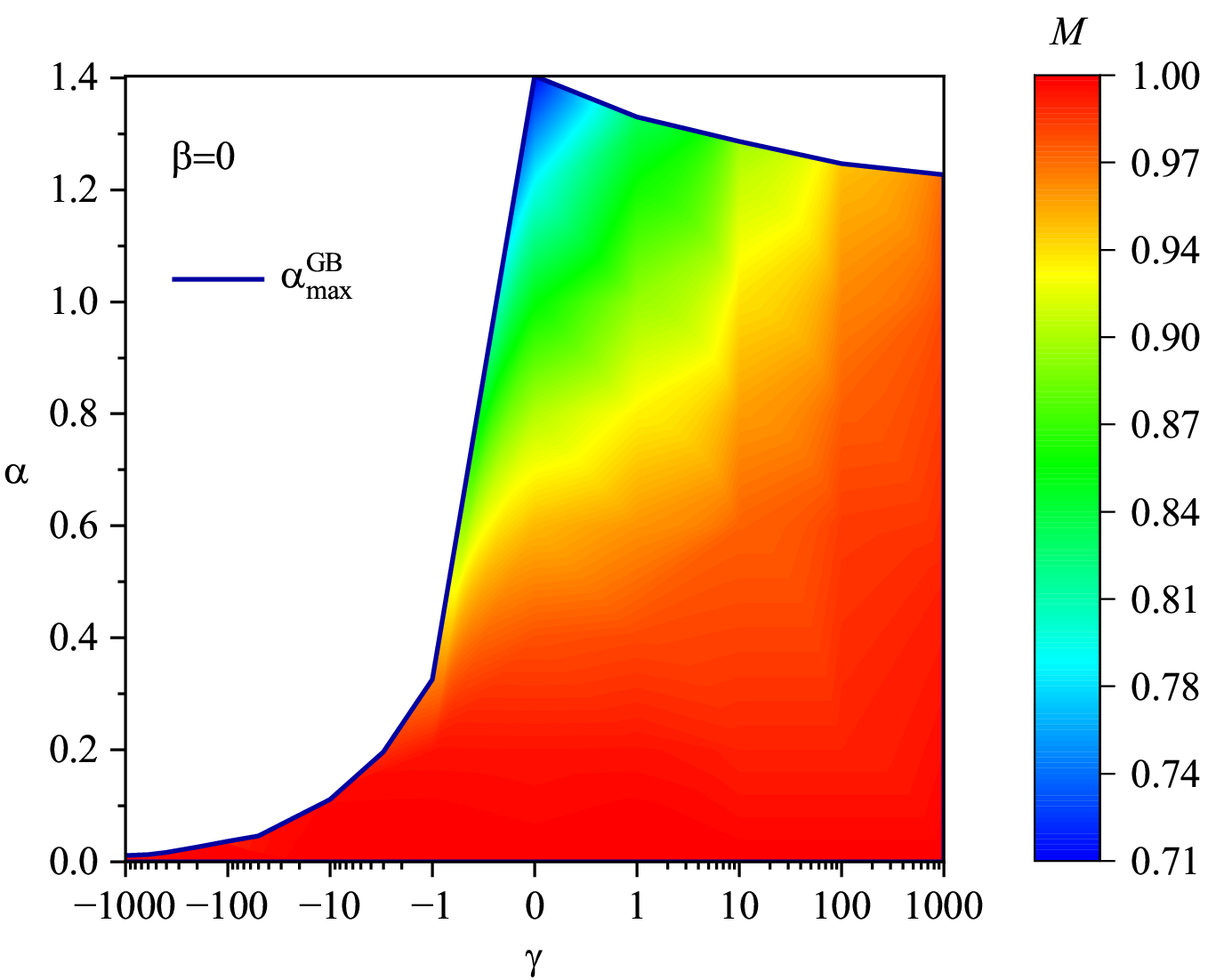}
        \end{center}
\end{minipage}\hfill
\begin{minipage}[t]{.49\linewidth}
        \begin{center}
\includegraphics[width=1.\linewidth]{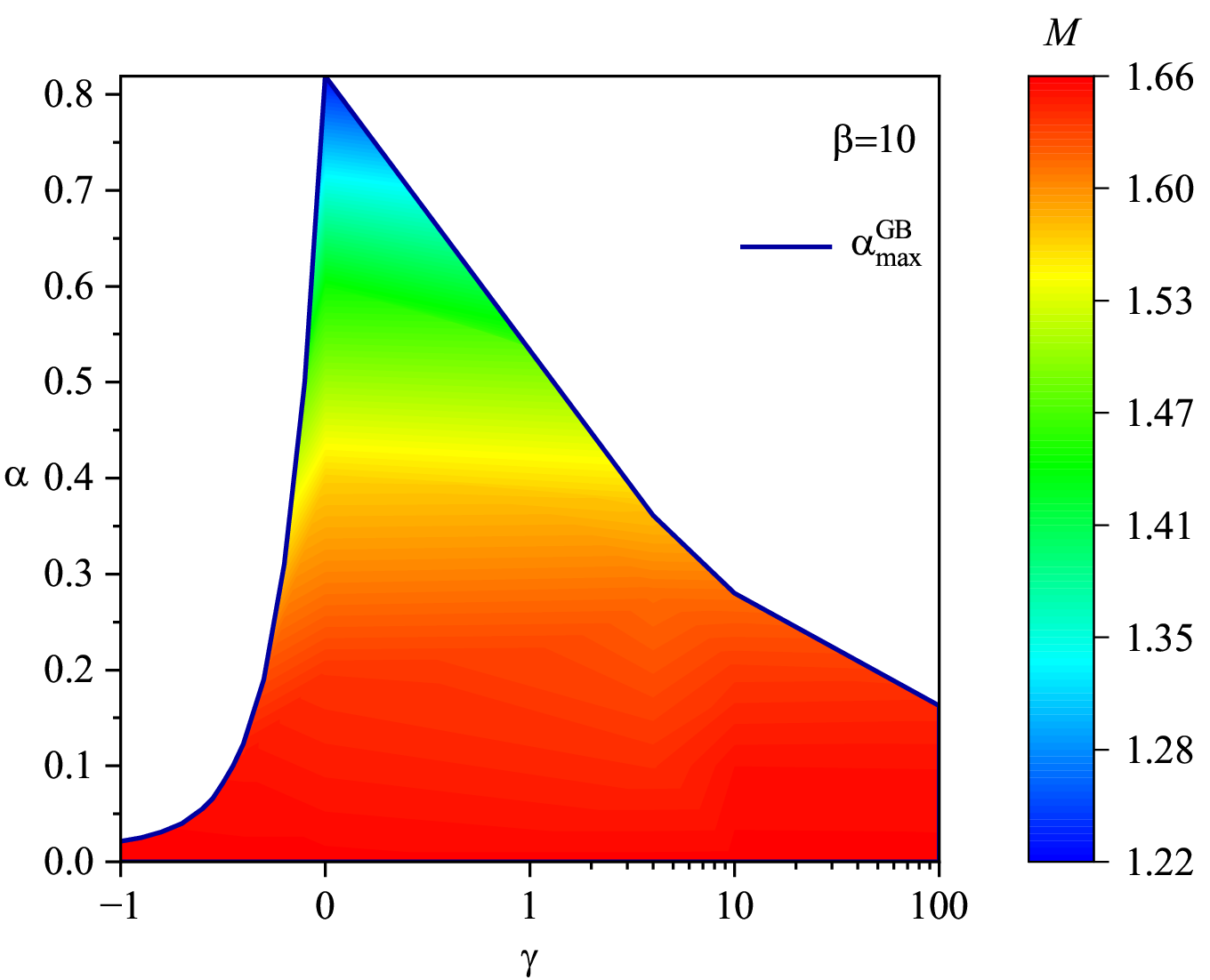}
        \end{center}
\end{minipage}\hfill
\begin{minipage}[t]{.49\linewidth}
        \begin{center}
\includegraphics[width=1.\linewidth]{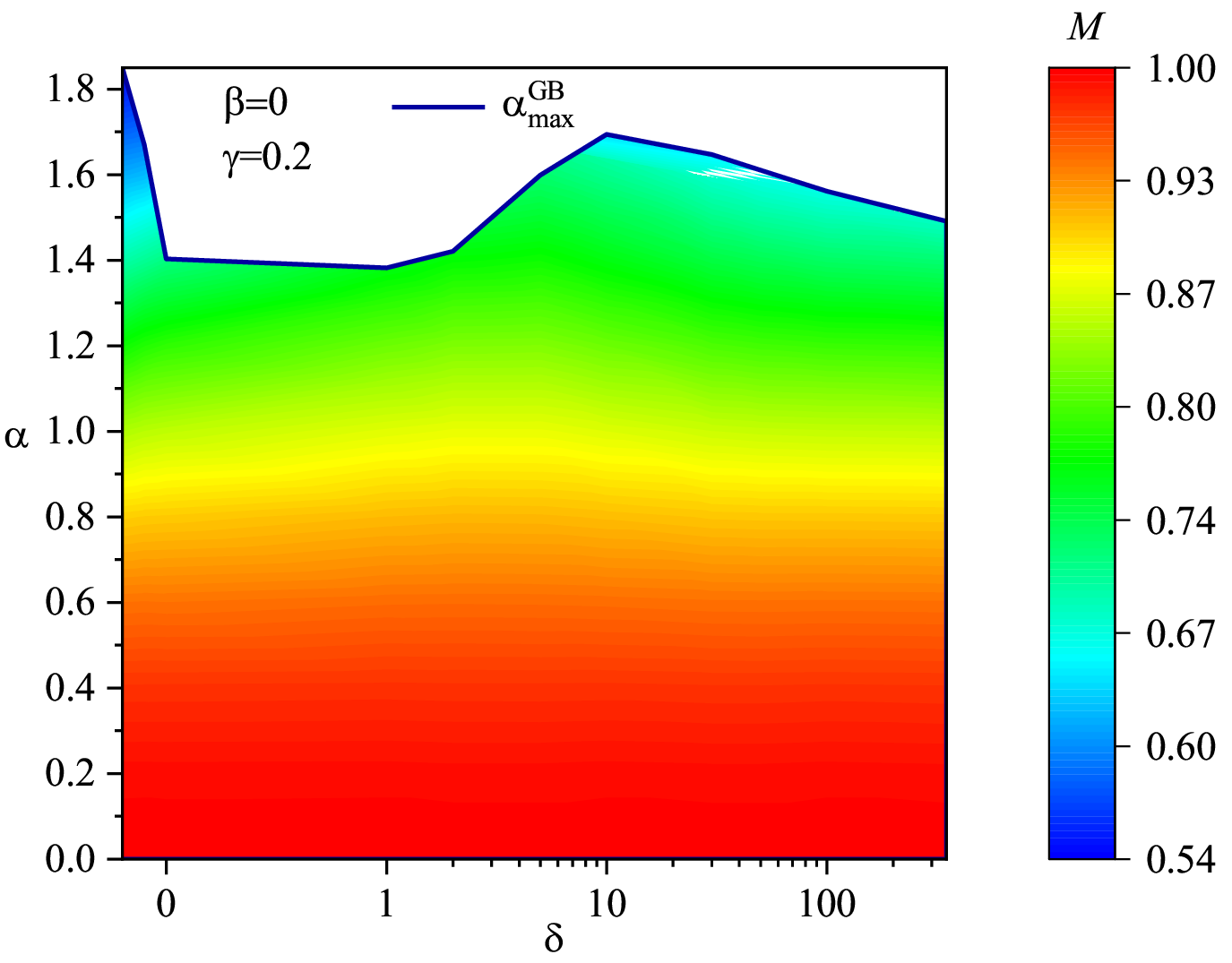}
        \end{center}
\end{minipage}\hfill
\begin{minipage}[t]{.49\linewidth}
        \begin{center}
\includegraphics[width=1.\linewidth]{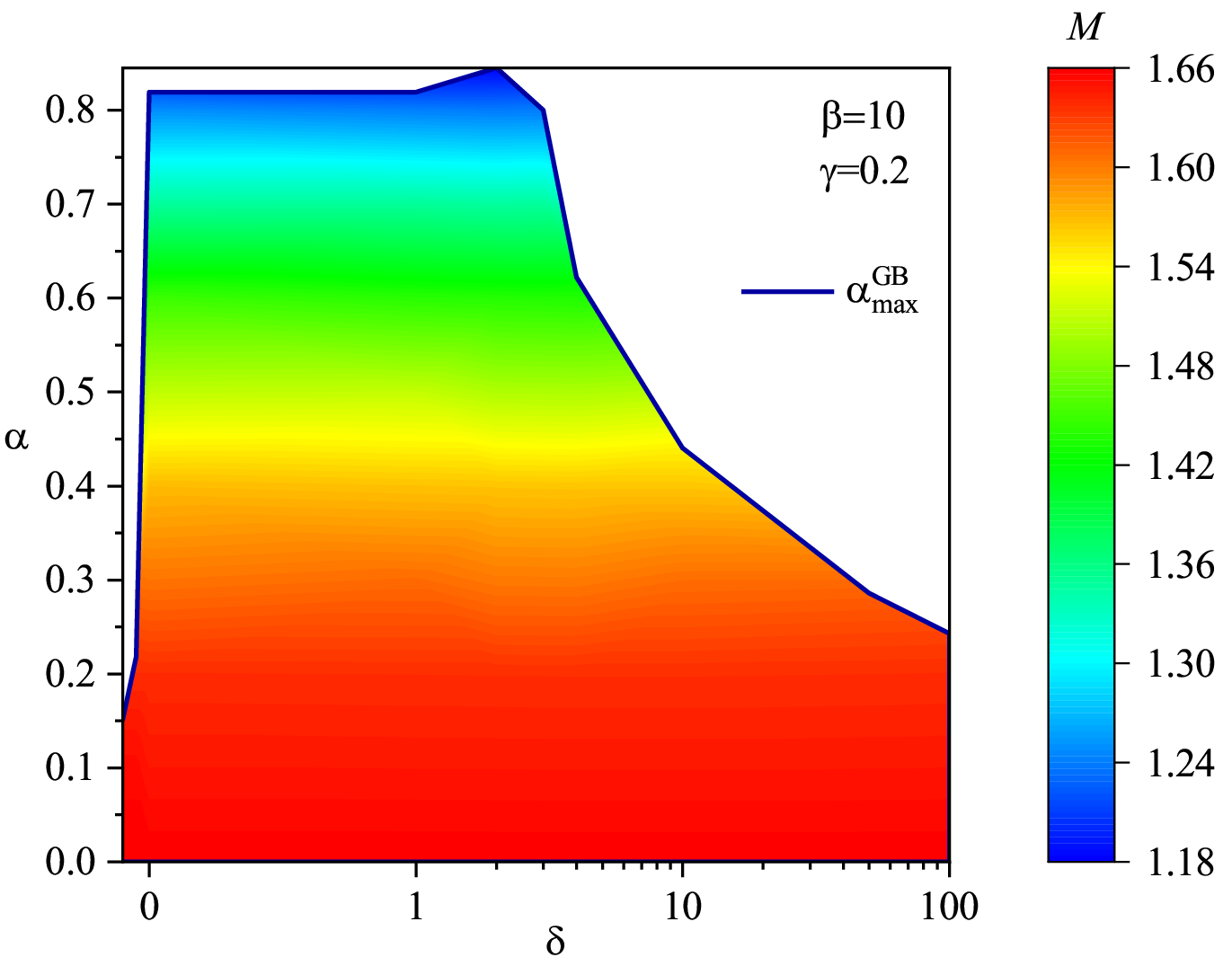}
        \end{center}
\end{minipage}\hfill
\caption{Domains of existence of the solutions [restricted by the lines $\alpha_{\text{max}}^{\text{GB}}(\gamma)$ and $\alpha_{\text{max}}^{\text{GB}}(\delta)$]
and distributions of the dimensionless ADM mass  from Eq.~\eqref{expres_mass} as functions of the effective
gravitational coupling constant $\alpha$ and the coupling constants $\gamma$ and $\delta$ from Eqs.~\eqref{pow_coup} (upper panels) and~\eqref{exp_coup} (lower panels)
are shown for $\beta=0, 10$. For all cases, the mass is given for configurations belonging to stable branches.
}
\label{fig_mass_domain}
\end{figure}

In this subsection, we investigate the dependence of the gravitating monopole solutions on the parameters
$\alpha, \beta$,  $\gamma$, and $\delta$. Before proceeding, we recall that
in GR ($\gamma,\delta=0$), when $\alpha$ increases from zero while $\beta$ is kept fixed, the fundamental branches of gravitating monopole solutions 
emerge smoothly from the corresponding flat-space monopole solutions~\cite{Breitenlohner:1991aa,Hartmann:2001ic}. 
Depending on the value of the Higgs self-coupling $\beta$,  there  exists a maximum value  $\alpha_{\text{max}}^{\text{GR}}(\beta)$  beyond which solutions no longer exist.
However, in the Bogomolnyi-Prasad-Sommerfield (BPS) limit  ($\beta=0$) and for small $\beta$, there is 
 a short backward branch of unstable solutions. This branch extends down to a critical value
 $\alpha_{\text{cr}}^{\text{GR}}(\beta)<\alpha_{\text{max}}^{\text{GR}}(\beta)$, where the monopole branch reaches 
 a limiting solution and bifurcates with the branch of extremal Reissner-Nordstr\"{o}m (RN) black-hole solutions~\cite{Breitenlohner:1991aa}.
For  larger $\beta$, such a backward branch is  absent, and $\alpha_{\text{max}}^{\text{GR}}(\beta)$  coincides with $\alpha_{\text{cr}}^{\text{GR}}(\beta)$.

For GB gravity, the situation with the possible behavior of solutions is more complex, and it is largely determined by the specific type of the coupling function.

\subsubsection{The case of the coupling function \eqref{pow_coup}}

As noted above, a characteristic feature of monopoles in GR is that, for any value of $\beta$, the limiting solution as $\alpha\to\alpha_{\text{cr}}^{\text{GR}}$ 
is always the extremal RN solution. For GB gravity with the coupling function \eqref{pow_coup}, a backward branch of unstable solutions also 
emerges for positive values of $\gamma \ll 1$, but this branch is significantly longer than in GR (see the left panel of Fig.~\ref{fig_mass_alpha},  the case $\gamma=0.2$). 
As in GR, this branch extends down to a critical value $\alpha_{\text{cr}}^{\text{GB}}<\alpha_{\text{max}}^{\text{GB}}$, 
where it eventually bifurcates with the branch of extremal RN black-hole solutions.

For $\gamma\gg 1$, we obtain solutions up to a certain value $\alpha=\alpha_{\text{max}}^{\text{GB}}$, 
while the corresponding configurations remain physically regular (see the left panel of Fig.~\ref{fig_mass_alpha},  the cases $\gamma=10$ and $\gamma=100$). 
However, as $\alpha\to\alpha_{\text{max}}^{\text{GB}}$, the solutions develop a cusp singularity~\cite{Kleihaus:2019rbg}. 
This occurs because the principal-part matrix $A$, whose entries are the coefficients multiplying the highest derivatives in the set of equations~\eqref{YM_eqs}--\eqref{Lambda_eqn}, 
becomes degenerate: its determinant vanishes locally at a critical point $x=x_{\text{max}}$. As a consequence, the first and second derivatives develop jump discontinuities at $x_{\text{max}}$. 
Fig.~\ref{fig_sols} illustrates the development of the cusp singularity as $\alpha$ increases for the system with $\gamma=10$ and $\beta=0$, 
where the singularity first appears at $\alpha_{\text{max}}^{\text{GB}}\approx1.28667$.

Unfortunately, the numerical method employed in this work does not allow us to identify all solution branches for intermediate values of $\gamma\sim 1$. 
Using the present numerical method, we find only the stable branches that emerge from the flat-space limit at $\alpha=0$ and extend up to a value $\alpha=\alpha_{\text{max}}^{\text{GB}}$, 
where the determinant of $A$ remains far from zero and the corresponding solutions are still far from the black-hole limit. 
Recovering the missing branches will likely require more advanced numerical techniques, such as pseudo-arclength continuation, which would enable continuation onto the unstable branches.

Returning to the case $\gamma \gg 1$, the situation becomes qualitatively different. In GR, the limitation on the existence of regular monopole solutions is global: 
it arises when the geometric size of the monopole core becomes comparable to its gravitational radius, leading to the formation of an event horizon. 
In EsGB gravity, by contrast, the limitation is local and develops inside the monopole core as a consequence of the nonlinear interaction 
between the curvature invariants and the scalar-field gradients. As the coupling constant $\alpha$ increases, the nonlinear contributions from the first derivatives of the Higgs 
and gauge fields modify the principal-part matrix $A$, and at $\alpha=\alpha_{\text{max}}^{\text{GB}}$
 its determinant vanishes. At this point the system loses ellipticity. Physically, this indicates that a static balance between the gauge-field pressure and the Gauss--Bonnet--induced 
 gravitational interaction can no longer be maintained. As shown in Fig.~\ref{fig_sols}, this manifests itself through the formation of a cusp singularity, 
 characterized by jump discontinuities in the first and second derivatives of the metric functions and in the second derivatives of the matter fields, 
 signaling a continuous geometric phase transition. Since no regular solutions of the ODE set exist for $\alpha>\alpha_{\text{max}}^{\text{GB}}$, 
 the monopole can no longer evolve toward the black-hole limit, as it does in GR or for small values of $\gamma$. Instead, strong Gauss--Bonnet effects preclude 
 the existence of regular monopole solutions long before the spacetime curvature becomes arbitrarily large.

The mathematical origin of the termination of the regular monopole branch described above suggests the following physical picture. 
As the coupling constant $\alpha $ increases, the nonlinear curvature invariant $R_{\text{GB}}^{2}$ in the core region undergoes a rapid nonlinear growth
(see Fig.~\ref{fig_sols}). Since in Gauss--Bonnet theory the geometry effectively interacts with field gradients through the field equations, 
the high energy density of the monopole in the core starts to 'induce' an effective mass of the fields. At the same time, a region forms inside the core 
where the geometric Gauss--Bonnet contribution to the energy-momentum tensor becomes locally dominant over the contribution from the Higgs field. 
As $\alpha \to \alpha_{\text{max}}^{\text{GB}}$, there is a point $x=x_{\text{max}}$ (cf. Fig.~\ref{fig_sols}), 
which acts as the macroscopic interface between two geometric phases: (i)~The inner phase ($x < x_{\text{max}}$) corresponds to the region 
where the effect of modified gravity is most significant (a possible physical interpretation of this situation is discussed at the end of Sec.~\ref{concl}). 
Here, the invariant $R_{\text{GB}}^{2}$ changes the sign of the Gauss--Bonnet term in the field equations, preventing the gravitational contraction of the core.
(ii)~The outer phase ($x > x_{\text{max}}$) is the region of standard GR asymptotics. The discontinuity in the second derivatives $H^{\prime\prime}, W^{\prime\prime}, \nu^{\prime\prime}$ 
(with continuous first derivatives of these functions) and in the invariant $R_{\text{GB}}^{2}$ itself at the point $x=x_{\text{max}}$ may be interpreted as a second-order (or higher-order) 
phase transition occurring at this point. At this point, the spatial structure of the monopole undergoes an abrupt rearrangement: the smooth transition between the monopole core 
and the outer vacuum is replaced by a thin-shell-like interface. For $\alpha > \alpha_{\text{max}}^{\text{GB}}$, the energy located in the core can no longer be supported by a smooth static field configuration.

The mathematical vanishing of the determinant of the principal-part matrix $A$ suggests the following analogy with thermodynamic phase transitions. 
Within this analogy, it may be viewed as corresponding to a diverging susceptibility accompanied by the critical slowing down of fluctuations. 
Motivated by this interpretation, one may speculate that, for $\alpha > \alpha_{\text{max}}^{\text{GB}}$, 
the monopole configuration effectively loses its ability to respond elastically: the gauge and Higgs fields can no longer compensate 
for the local pressure generated by the Gauss--Bonnet--induced curvature effects. In this picture, the static monopole phase becomes thermodynamically unstable and ceases to exist.

In light of the above, the left panel of Fig.~\ref{fig_mass_alpha} illustrates the typical behavior of the mass curves for various values 
of the parameter $\gamma$ in the BPS limit. Fig.~\ref{fig_mass_alpha} shows that the value of $\alpha_{\text{max}}^{\text{GB}}$ 
is always smaller than the corresponding value in GR and decreases with increasing $\gamma$. The situation is similar for negative values of $\gamma$, 
where $\alpha_{\text{max}}^{\text{GB}}$ decreases rapidly with increasing $|\gamma|$ (cf. Fig.~\ref{fig_mass_domain}). 
At a fixed gravitational coupling constant $\alpha$, the monopole mass in GB gravity is always larger for positive $\gamma$ and smaller 
for negative $\gamma$ than that of the corresponding GR solutions.

The corresponding domains of existence are shown in the upper panels of Fig.~\ref{fig_mass_domain}, where the total monopole mass is plotted 
for various values of the parameters $\alpha$, $\beta$, and $\gamma$ for which we could obtain solutions using the present numerical method. 
The curve $\alpha_{\text{max}}^{\text{GB}}(\gamma)$ forms 
the boundary of the domain of existence of stable solutions. Regular solutions exist for both positive and negative values of $\gamma$. 
Moreover, the domain of existence is highly asymmetric with respect to the GR case ($\gamma=0$), both in the BPS limit and for $\beta\neq 0$. 
In particular, in the BPS limit, $\alpha_{\text{max}}^{\text{GB}}$ depends only weakly on positive values of $\gamma$, whereas for small negative 
values of $\gamma$ it drops sharply, followed by a gradual decrease toward values close to zero. For $\beta=10$, by contrast, $\alpha_{\text{max}}^{\text{GB}}$ 
decreases much more rapidly with both positive and negative values of $\gamma$ than in the BPS limit. We also note that the GR solutions 
have the lowest monopole mass among all the cases considered.

\subsubsection{The case of the coupling function \eqref{exp_coup}}

In this case, $\alpha _{\text{max}}^{\text{GB}}$ can be either larger (for negative values of $\delta $) or smaller (for positive values of $\delta $) than the corresponding GR value. 
To illustrate the effect of the Gauss--Bonnet modification, we compare the GR solutions ($\delta=0$)  with the EsGB solutions for $\gamma=0.2$ over the full 
range of $\alpha $ for three representative values of the parameter, $\delta=-0.2$, $\delta=1$, and $\delta=100$ (see the right panel of Fig.~\ref{fig_mass_alpha}). 
The figure shows that, in the BPS limit, the mass curves differ noticeably between the GR and EsGB cases. For $\delta=-0.2$, the minimum mass attained as $\alpha \to \alpha_{\text{max}}^{\text{GB}}$ 
is significantly smaller than the corresponding GR value, and, as in GR when $\alpha \to \alpha_{\text{cr}}^{\text{GR}}$, 
the solution approaches the extremal RN branch. For $\delta=1$, by contrast, the minimum mass is larger than in GR. In this case, 
as in GR and for the coupling function \eqref{pow_coup}, a backward branch of unstable solutions bifurcates from $\alpha \to \alpha_{\text{max}}^{\text{GB}}$, 
although it is considerably longer than the corresponding GR branch. We also note that the configurations obtained for $\delta=1$ are close 
to those found for the coupling function \eqref{pow_coup} with $\gamma=0.2$ (cf. the corresponding curves in Fig.~\ref{fig_mass_alpha}). Finally, for $\delta=100$, the stable branch terminates at 
$\alpha \to \alpha_{\text{max}}^{\text{GB}}\approx\alpha_{\text{cr}}^{\text{GB}}$, and the limiting configuration appears to approach a hairy (non-RN) black-hole solution.

Obtaining solutions for $\delta>1$ is numerically challenging. The numerical method employed in this work yields the following results (both in the BPS limit and for $\beta\neq 0$). 
For $1\lesssim \delta \lesssim 10$, solutions can be obtained up to maximum values $\alpha _{\text{max}}^{\text{GB}}$, 
beyond which the present numerical method no longer converges. At these values of $\alpha _{\text{max}}^{\text{GB}}$, 
the determinant of $A$ remains far from zero, and the corresponding solutions are still regular rather than black-hole configurations. 
By analogy with GR and with the polynomial coupling~\eqref{pow_coup}, it is plausible that the stable branch bifurcates with an unstable branch at this point. 
For $10\lesssim \delta \lesssim 100$, we observe two distinct types of behavior. (i)~For $\beta=0$, with increasing $\delta$, the limiting solutions gradually approach hairy black-hole configurations. 
(ii) For $\beta=10$, solutions terminate at a value of $\alpha _{\text{max}}^{\text{GB}}$ 
 while remaining far from the black-hole limit. Finally, for even larger values of $\delta $ and $\beta=0$, the limiting solutions again approach the branch of extremal RN black-hole solutions 
 (i.e., $\alpha_{\text{max}}^{\text{GB}}\to \alpha_{\text{cr}}^{\text{GR}}$), as expected, since for very large $\delta $ the exponential term in Eq.~\eqref{exp_coup} becomes negligible 
 and the EsGB solutions reduce to their GR counterparts. For $\beta= 10$, however, increasing $\delta $ leads to progressively shorter stable branches that do not approach the black-hole limit.

The corresponding domains of existence are shown in the lower panels of Fig.~\ref{fig_mass_domain} for various values of the parameters $\alpha, \beta$, and $\delta$. 
Regular solutions exist for both positive and negative values of $\delta $. Moreover, as for the polynomial coupling~\eqref{pow_coup}, 
the domain of existence is highly asymmetric with respect to the GR case ($\delta=0$). However, the qualitative dependence of $\alpha_{\text{max}}^{\text{GB}}$
 on $\delta$ differs significantly from that found for the polynomial coupling, as is evident from the corresponding plots. For the polynomial coupling, the maximum of the curve $\alpha_{\text{max}}^{\text{GB}}(\gamma)$  
 is always located at $\gamma=0$, i.e., in the GR case. By contrast, for the exponential coupling, the maxima of $\alpha_{\text{max}}^{\text{GB}}(\delta)$ 
 occur for nonzero values of $\delta $: in the BPS limit there are two such maxima, corresponding to positive and negative values of $\delta $, 
 whereas for $\beta=10$ only one maximum remains, located at a positive value of $\delta $. We also note that, unlike the polynomial coupling, 
 the minimum monopole mass is attained in GB gravity rather than in GR~-- in the BPS limit at a certain negative value of $\delta $, and for $\beta=10$ at a certain positive value of  $\delta $.

 \section{Conclusions and discussion}
\label{concl}

In this paper, we have studied self-gravitating 't Hooft--Polyakov monopoles in Einstein--scalar--Gauss--Bonnet gravity. 
Our analysis has focused on static, spherically symmetric configurations corresponding to the fundamental (nodeless) branch of solutions. 
We have considered two classes of scalar--Gauss--Bonnet coupling functions, polynomial and exponential, and have shown 
that the resulting phase structure depends sensitively on both the choice of the coupling function and the values of its parameters.

Our analysis shows that, unlike GR, where the solution branch eventually and inevitably approaches the extremal RN black hole, 
in EsGB gravity there are several distinct branch-termination scenarios depending on the form of the coupling function. Specifically,

\begin{itemize}
\item In the case of the polynomial coupling of type \eqref{pow_coup}, the following situation occurs. 
Unlike GR, where the existence limit for 't Hooft--Polyakov monopoles is determined by their gravitational collapse 
into an extremal RN black hole, in Gauss--Bonnet gravity a limit of this type is realized only for small $\gamma $, 
whereas for $|\gamma|\gg 1$ the limit of existence has a fundamentally different, nonperturbative nature. 
We have shown that the regular monopole solution branch terminates long before the formation of an event horizon 
due to the vanishing of the determinant of the principal-part matrix $A$. This degeneracy of the principal part of the equations 
leads to a cusp in the metric components and, as a result, to a geometric phase transition at strictly finite values of the metric functions, 
making the classical RN limit physically unattainable.
\item For the exponential coupling~\eqref{exp_coup}, the branch structure is considerably richer and is primarily controlled by the parameter $\delta $.
\item[(i)] For small negative $\delta $, there are branches of regular solutions that lie below the corresponding GR branch and, just as in GR, 
approach the extremal RN black-hole solutions. In this case, the minimum masses of such monopoles can be significantly smaller than the monopole mass in GR.
\item[(ii)] For positive $\delta\lesssim 1$, the minimum monopole mass is reached as $\alpha \to \alpha_{\text{max}}^{\text{GB}}$, 
and it is larger than the monopole mass in GR. In this case, just as in GR, at the point $\alpha \to \alpha_{\text{max}}^{\text{GB}}$ 
a backward branch emerges which eventually bifurcates with the branch of extremal RN black-hole solutions, but this branch can be significantly longer than in GR.
\item[(iii)] For intermediate values of $1\lesssim \delta\lesssim 10$, regular solutions exist up to a certain maximum value $\alpha _{\text{max}}^{\text{GB}}$, 
 beyond which no further solutions could be obtained with the present numerical method. Here, the determinant of $A$ is still significantly different from zero, and the solution remains regular (a non-BH solution).
 \item[(iv)] For $10\lesssim \delta \lesssim 100$, the branch gradually evolves toward those values of $\alpha _{\text{max}}^{\text{GB}}$ 
 at which the solutions begin to approach either to hairy black-hole configurations  (for $\beta=0$) or to non-BH systems (for $\beta=10$).
 \item[(v)] For still larger values of $\delta $, in the BPS limit, the limiting solutions again tend to the branch of extremal RN black-hole solutions 
 (i.e., $\alpha_{\text{max}}^{\text{GB}}\to \alpha_{\text{cr}}^{\text{GR}}$): here, the exponential term in \eqref{exp_coup} becomes negligible, and the solutions approach the GR limit. 
 In turn, for $\beta= 10$, increasing $\delta $ results in progressively shorter branches of stable solutions that do not reach a black hole.
\end{itemize}

A comparison of the two coupling functions shows that the structure of the monopole solution branches depends critically 
on the functional form of the scalar--Gauss--Bonnet coupling. For the polynomial coupling with $|\gamma|\gg 1$, 
the existence limit of static monopoles has a genuinely nonperturbative geometric origin: the system undergoes a local 
degeneration of the principal part of the field equations ($\det A \to 0$), leading to a continuous (second- or higher-order) geometric 
phase transition with a cusp in the metric functions while the spacetime curvature remains finite. By contrast, the exponential coupling 
acts as a natural regulator that prevents the degeneration of the principal part ($\det A \neq 0$) and preserves the smoothness of the field profiles. 
In this case, the stability limit of the monopole is instead determined 
by a classical macroscopic branch-merging bifurcation (turning point). Thus, whereas the polynomial coupling leads to a local 
breakdown of the principal part of the field equations, the exponential coupling may presumably permit additional unstable branches 
that approach hairy black-hole configurations.

The numerical method used in the present work allowed us to identify only the solution branches discussed above. 
This, however, does not imply that additional branches of fundamental (nodeless) solutions, whether stable or unstable, do not exist. 
Identifying such branches will likely require alternative numerical techniques, such as pseudo-arclength continuation. 
Such an approach would enable a more complete analysis of the bifurcation structure of the solutions. 
We plan to pursue this in future work in order to construct a more complete phase diagram of the monopole solutions in the parameter space. 
This should clarify how the choice of the coupling function $f(\Phi)$  determines not only the quantitative properties of the monopoles 
but also the qualitative mechanism governing the termination of the regular solution branches.

The present results demonstrate that the nonlinear dynamics of monopole solutions in EsGB gravity differs qualitatively from that in general relativity. 
Since GR is a well-established classical theory of gravity, an important question is under what conditions EsGB gravity may provide a realistic description of physical systems? 
One possible interpretation is that modified theories of gravity, including EsGB gravity, may be regarded as effective theories incorporating quantum-gravitational corrections 
(see, e.g., Refs.~\cite{Dzhunushaliev:2013nea,Dzhunushaliev:2015mva}). Within this interpretation, the critical solutions with cusp singularities found here and, for example, in Ref.~\cite{Kleihaus:2019rbg} 
may suggest that quantum-gravitational effects become dominant in the inner region ($x < x_{\text{max}}$), where they are approximately described by the Gauss--Bonnet corrections, 
whereas in the outer region ($x > x_{\text{max}}$) the spacetime dynamics is well described by classical GR.
While this interpretation is necessarily speculative, it provides a possible physical meaning of the critical solutions obtained in the present work.

\end{document}